\documentclass[numreferences]{kluwer}    

\begin{document}                                                                                   
\begin{article}
\begin{opening}         
\title{Gravity from Spacetime Thermodynamics} 
\author{T. Padmanabhan}  
\runningauthor{T. Padmanabhan}
\runningtitle{Gravity from spacetime thermodynamics}
\institute{IUCAA, Pune University Campus, Pune 411 007\\
\email{nabhan@iucaa.ernet.in}}

\begin{abstract}
The Einstein-Hilbert action (and thus  the dynamics of gravity) can be obtained
by: (i)  combining the 
principle of equivalence,
    special relativity  and quantum theory in the 
    Rindler frame and (ii) postulating that
the horizon area must be proportional to the entropy. This approach  uses the 
 local Rindler frame  as a natural extension of the 
local inertial frame, and 
 leads to the interpretation that the gravitational action represents the free energy of the 
spacetime geometry.
 As an aside, one  obtains  an insight into the peculiar structure of Einstein-Hilbert action and a natural explanation  to the questions:
(i) Why does the covariant action for gravity contain second derivatives of the metric tensor? 
(ii)  Why is the   gravitational coupling constant is positive ?
Some geometrical features of gravitational action are clarified.
\end{abstract}

\end{opening}

    \section{Introduction  and summary}
    
     The (i) existence of the principle of equivalence and (ii) 
     the connection between gravity and thermodynamics
     are the two most surprising features of gravity. Among these two, the
      principle of equivalence  finds its natural expression when gravity is described as a manifestation of
     curved spacetime. 
     This --- in turn --- makes gravity  the only interaction which is capable
     of wrapping up regions of spacetime so that information from one region is not 
     accessible to observers at another region. Given the fact that entropy of a system
     is closely related to accessibility of information, it is inevitable that there will be
    some connection between gravity and thermodynamics (for a review, see references \cite{birreltp}
    \cite{tprealms}). 
    But, in contrast to the principle of equivalence,  years of 
    research in this field (see, for a sample of references 
    \cite{bhentropy}),  
    has not 
  led to  something more profound or fundamental arising out of this feature. 
    
   This suggests that we should  learn a lesson from the way Einstein handled the principle of 
    equivalence and  apply it in the context of the connection between thermodynamics
    and gravity. Einstein did not attempt to ``derive" principle of equivalence in the conventional 
    sense of the word. Rather, he accepted it as a key feature which must find expression in the 
    way gravity is described --- thereby obtaining  a geometrical description of gravity. 
    Once the geometrical interpretation of gravity is accepted, it follows that there {\it will} arise
    surfaces which act as one-way-membranes for information and  will thus lead to some connection
    with thermodynamics. It is, therefore, more in tune with the spirit of Einstein's analysis
    to {\it accept} an inevitable connection between gravity and thermodynamics and ask 
    what such a connection would imply. 
   I will now elaborate this idea further in order to show how powerful it is \cite{grf}. 
    
    The first step in the logic,  the principle of equivalence, 
     allows one to define a coordinate system around any event ${\cal P}$ in a region of size
    $L$ (with $L^2(\partial^2 g/g)   \ll 1$ but $L(\partial g/g) $ being arbitrary)
    in  which the spacetime is locally inertial.   
    As the second step, we want to give expression to the fact that there is a deep connection between
    one-way-membranes arising in a spacetime and thermodynamical entropy. This, of course,
    is not possible in the local inertial frame since the quantum field theory in that frame, say, 
    does not recognize any non trivial geometry of spacetime. But it is possible to achieve our aim
    by using a uniformly accelerated frame around ${\cal P}$. In fact, 
    around  any event ${\cal P}$ we have fiducial observers anchored
    firmly   in space with ${\bf x} = $ constant and  the four-velocity $u^i = g_{00}^{-1/2}(1,0,0,0)$
    and  acceleration $a^i = u^j \nabla_j u^i$. This allows us to define a second natural
    coordinate system around any event by using the Fermi-Walker transported coordinates
    corresponding to these accelerated observers. I shall call this the local Rindler frame.
    [Operationally, this coordinate system is most easily constructed by first transforming to
    the locally inertial frame and then using the standard transformations between the 
    inertial coordinates and the Rindler coordinates.]
    This local Rindler frame will lead to a natural notion of horizon and associated temperature.
    The key new idea  will be to postulate that the horizon in the 
    local Rindler frame also has an entropy per unit transverse area and  demand that any
    description of gravity must have this feature incorporated in it. 
    
    What will such a postulate lead to? Incredibly enough, it leads to the {\it correct Einstein-Hilbert
    action principle}   for gravity.
     Note that the original approach of Einstein making use of the principle of
    equivalence lead only up to the {\it kinematics} of gravity --- viz., that gravity is described by a curved
    spacetime with a non trivial metric $g_{ab}$ --- and cannot tell us how the {\it dynamics} of the 
    spacetime is determined. Taking the next step, using the local Rindler frame and demanding that
    gravity must incorporate the thermodynamical aspects lead to the action functional itself.

    This  approach also throws light
    on (what has been usually considered) a completely different issue: Why does
    the Einstein-Hilbert action contain second derivatives of the metric tensor?
     The new approach ``builds up''
    the Einstein-Hilbert action from its surface behaviour and, in this sense,
    shows that gravity is intrinsically holographic \cite{holo}. 
     I use this term with the specific meaning
    that given the form of the action on a two dimensional surface, there is a way of obtaining
    the full bulk action. 
In the $(3+1)$ formalism, this leads to the interpretation of the gravitational action as the
free energy of spacetime.  Einstein's equations are equivalent to the principle of minimization of 
free energy in thermodynamics.

    \section{Gravitational dynamics from spacetime thermodynamics}

    The principle of equivalence leads to a geometrical description of gravity in which
    $g_{ab}$ are the fundamental variables. So we expect the dynamics of gravity to be described by some
    {\it unknown} action functional 
    \begin{equation}
    A =
     \int d^4x \sqrt{-g} L(g,\partial g) \equiv \int d^4 x \sqrt{-g} L(g, \Gamma)
    \label{aquad}
    \end{equation}
    involving $g_{ab}$s and their first derivatives $\partial_c g_{ab}$ or, 
    equivalently, the set $[g_{ab}, \Gamma^i_{jk}]$ where $\Gamma$s are the standard 
    Christoffel symbols.

    Given any Lagrangian $L(\partial q, q)$ involving only up to the first derivatives of the
     dynamical variables, it is 
         {\it always} possible
to construct another Lagrangian $L'(\partial^2q,\partial q,q)$, involving 
second derivatives such that it describes the same dynamics \cite{tpdlb}.  This idea  works for any number of variables
    (so that $q $ can be a multicomponent entity) dependent on space and time.
    But I shall illustrate it in the context of point mechanics. The prescription is:
\begin{equation}
L'=L-{d\over dt}\left(q{\partial L\over \partial\dot q}\right)
\label{lbtp}
\end{equation}
While varying the $L'$, one keeps the {\it momenta} $(\partial L/\partial\dot q)$ fixed
at the endpoints rather than $q'$s. 
This is most easily seen by explicit variation; we have
   \begin{eqnarray}
 \delta A' &= &    \int\limits^{{\cal P}_2}_{{\cal P}_1} dt \left[ {\partial L \over \partial  q }\delta q +  {\partial L \over \partial \dot q }\delta \dot q \right] - \delta \left(  q {\partial L \over \partial \dot q } \right) \bigg \vert^{{\cal P}_2}_{{\cal P}_1}\nonumber \\
&= &\int\limits^{{\cal P}_2}_{{\cal P}_1}  dt \left[ {\partial L \over \partial  q } - {d \over dt} \left(  {\partial L \over \partial \dot q } \right) \right] \delta q - q \delta p \bigg \vert^{{\cal P}_2}_{{\cal P}_1}
\end{eqnarray}
If we keep $\delta p = 0$ at the end points while varying $L'$, then we get back the same
Euler-Lagrange equations  as obtained by varying $L$ and keeping $\delta q =0$ at end points.
Since $L = L ( \dot q, q )$, the quantity $q ( \partial L /\partial \dot q )$ will also depend on $\dot q $ and
 the term $d(q \partial L / \partial \dot q) / dt $ will involve $\ddot q$. Thus $L'$ contains second derivatives of $q$ while $L$ contains only up to first derivatives. In spite of the fact that $L'$ contains second derivatives of $q$, the equations of motion arising from $L'$ are only second order for variation
with $\delta p =0$ at end points.
It can be shown that, in the  path integral formulation of quantum theory,  the modified Lagrangian $L'$ 
correctly describes the transition amplitude between states with given momenta
(see p. 170 of  \cite{tpdlb}). 
    
  Thus, in the case of gravity,  the {\it same}  equations
    of motion can be obtained from another (as yet unknown) action:
    \begin{eqnarray}
    A' &=& \int d^4x \sqrt{-g} L - \int d^4x \partial_c \left[ g_{ab}
     {\partial \sqrt{-g} L \over \partial(\partial_c g_{ab})}
    \right]  \nonumber \\
   &\equiv&  A - \int d^4 x \partial_c (\sqrt{-g}V^c) \equiv A - \int d^4 x \partial_c P^c 
    \label{aeh}
    \end{eqnarray}   
    where $V^c $ is made of $g_{ab} $ and $\Gamma^i_{jk}$. Further, $V^c$ must be linear
    in the $\Gamma$'s since the original Lagrangian $L$ was quadratic in the first derivatives
    of the metric. 
    Since $\Gamma$s vanish in the local inertial frame and the metric reduces to
    the Lorentzian form, the action $A$ cannot be generally covariant.
    However, the action $A'$ involves the second derivatives  of the metric and 
    we shall see later that that the action $A'$ is indeed generally covariant.

    To obtain a quantity $V^c$, which is linear in $\Gamma$s
     and having  a single index $c$, from $g_{ab} $ and $\Gamma^i_{jk}$,
     we must contract on two of the indices on $\Gamma$
    using the metric tensor. 
    (Note that we require $A$, $A'$ etc. to be Lorentz scalars and $P^c, V^c$ etc.
    to be vectors under Lorentz transformation.)
    Hence the most general choice for $V^c$ is the linear combination
      \begin{equation}
    V^c =  \left(a_1 g^{ck} \Gamma^m_{km} +a_2 g^{ik} \Gamma^c_{ik}\right) 
    \label{defvc}
     \end{equation}
     where $a_1(g)$ and $a_2(g)$ are unknown functions of the determinant $g$ of the metric
(which is the only (pseudo) scalar entity which can be constructed from $g_{ab}$s and $\Gamma^i_{jk}$s).
      Using the identities $\Gamma^m_{km} =\partial_k
     (\ln \sqrt{-g})$, \ $\sqrt{-g}g^{ik}\Gamma^c_{ik} = -\partial_b(\sqrt{-g}g^{bc})$,
     we can rewrite the expression for $P^c \equiv \sqrt{-g}V^c$ as 
     \begin{equation}
    P^c =\sqrt{-g}V^c=
    c_1(g)g^{cb} \partial_b \sqrt{-g} +c_2(g) \sqrt{-g} \partial_b g^{bc}
    \label{defpc}
    \end{equation}
    where $c_1\equiv a_1 - a_2,\  c_2\equiv -a_2$ are two other unknown functions of $g$. 
    If we can fix these coefficients by using a physically well motivated prescription, then
    we can determine the surface term and by integrating, the Lagrangian $L$.
    I will now show how this can be done.
    
    Let us consider a static spacetime in which all $g_{ab}$s are
    independent of $x^0$ and $g_{0\alpha} =0$. Around any given event ${\cal P}$
    one can construct a local Rindler frame with an acceleration 
    of the observers with ${\bf x} $ = constant, given by $a^i = (0,{\bf a})$ and
    ${\bf a}= \nabla (\ln \sqrt{g_{00}})$.
    This Rindler frame will have a horizon which is a plane surface normal to the 
    direction of acceleration and a temperature $T=|{\bf a}|/2\pi$ associated with this horizon.
     I shall postulate that the entropy associated with
    this horizon is proportional to its area or, more precisely, 
    \begin{equation}
    {dS\over dA_\perp} = {1\over {\cal A}_P}
    \label{postulate}
    \end{equation}
    where ${\cal A}_P$ is a fundamental constant with the dimensions of area. 
    It represents the minimum area required to hold unit amount of  information 
     and our postulate demands that this number be finite.
    Given the temperature of the horizon, one can construct a canonical ensemble 
    with this temperature and relate the Euclidean action to the thermodynamic entropy
   (see, e.g, \cite{tprealms} \cite{tplongpap}).
    Since the Euclidean action can be interpreted as the entropy in the canonical
    ensemble, I will demand that the surface term in equation (\ref{aeh})
    should be related to the entropy $S$ by  $S = -A_{\rm surface}$ 
    (with the minus sign arising from standard 
    Euclidean continuation \cite{tprealms}), when evaluated in the local
    Rindler frame with the temperature $T$. In particular, this result must hold in flat spacetime in Rindler coordinates.
    [We will see later that the action $A'$ is generally covariant and  will vanish
    in the  flat spacetime, in the absence of the cosmological constant. It follows that the numerical value of the action  $A$ in the 
    Rindler frame is the same as the surface term in equation (\ref{aeh}).]
     In the static Rindler frame, the surface term is
     \begin{equation}
     A_{\rm surface}  = \int d^4 x \partial_c P^c = \int_0^\beta dt \int_{\cal V} d^3 x
     \nabla \cdot {\bf P} = \beta \int_{\partial{\cal V}} d^2 x_\perp \hat{\bf n}\cdot {\bf P}
     \label{keyeqn}
     \end{equation}
     I  have restricted the time integration to an interval $(0,\beta)$
     where $\beta = (2\pi /|{\bf a}|) $ is the inverse temperature in the Rindler frame.
     This is {\it needed} since the Euclidean action will be periodic in the imaginary
     time with the period $\beta$. 
    We shall choose the  Rindler frame such that the acceleration is along
    the $x^1=x$ axis.
      The most general form of the metric representing the Rindler frame can be
  expressed in the form 
  \begin{eqnarray}
  ds^2 &=& (1+2al) dt^2 - \frac{dl^2}{(1+2al)} - (dy^2+dz^2)\nonumber \\
  &=& \left[ 1+2al(x) \right]dt^2 - \frac{l^{'2}}{[1+2al(x)]}dx^2 - (dy^2+dz^2)
  \label{rindmetric}
  \end{eqnarray}
  where $l(x)$ is an arbitrary function and $l' \equiv (dl/dx)$. [Since the acceleration is 
  along the x-axis,  the metric in the transverse direction is unaffected. The first form of the metric is the standard
Rindler frame in the $(t,l,y,z)$ coordinates. We can, however, make any coordinate transformation
from $l$ to some other variable $x$ without affecting the planar symmetry or the static nature of the metric. This leads to the general form of the metric given in the second line, in terms of the $(t,x,y,z)$ coordinates.]
  Evaluating the surface term $P^c$ in (\ref{defpc})
  for this metric, we get
  the only non zero component to be
  \begin{equation}
  P^x = -2ac_2(g) - [1+2al(x)] \frac{l''}{l^{'2}} [c_1(g)- 2 c_2(g)]
  \end{equation}
       so that the action in (\ref{keyeqn})
    becomes 
    \begin{equation}
    A = \beta P^x \int d^2x_\perp = \beta P^x A_\perp = -S
    \label{detcone}
    \end{equation}   
    where $A_\perp$ is the transverse area of the $(y-z)$ plane.
    The last equality identifies the entropy $S$, which is equal the Euclidean action, 
     with the minus sign arising from standard Euclidean continuation.
     From our postulate (\ref{postulate}) it follows that
     \begin{equation}
     \frac{dS}{dA_\perp} = 2 a \beta c_2 (g) + \beta [c_1 - 2 c_2] (1+2al) \frac{l''}{l^{'2}} = \frac{1}{{\cal A}_P}
     \label{xxx}
     \end{equation}
    For this quantity to be  a constant independent of $x$ for any choice of $l(x)$, 
  the second term must vanish
  requiring $c_1(g) = 2 c_2(g)$. 
  An explicit way of obtaining this result is to consider a class of functions $l(x)$ which satisfy the relation
  $l'=(1+2al)^n$ with $0\leq n\leq 1$. Then
  \begin{equation}
  \beta [c_1(l') - 2 c_2(l')] (1+2al) \frac{l''}{l^{'2}}=2a\beta[ c_1(l') - 2 c_2(l')]n
  \end{equation}
  which can be independent of $n$ and $x$ only if $c_1(g) = 2 c_2(g)$. 
  Further, using $a\beta = 2\pi$,
  we find that  $c_2(g)=(4\pi {\cal A}_p)^{-1}$
  which is a constant independent of $g$. 
    Hence $P^c$ has the form 
    \begin{eqnarray}
    P^c &=&   {1\over 4\pi {\cal A}_P} \left( 2g^{cb} \partial_b \sqrt{-g} + \sqrt{-g} \partial_b g^{bc}\right)
    ={\sqrt{-g}\over 4\pi {\cal A}_P}  \left( g^{ck} \Gamma^m_{km} - g^{ik} \Gamma^c_{ik}\right) \nonumber \\
    &=&- {1\over 4\pi {\cal A}_P}\frac{1}{\sqrt{-g}} \partial_b(gg^{bc})
    \label{pcfix}
    \end{eqnarray}
    The second equality is obtained by using the standard identities mentioned after 
    equation (\ref{defvc}) while the third equality follows directly by combining the 
    two terms in the first expression. 
    This result is remarkable and let me discuss it before proceeding further.
    
    The general form of $P^c$ which we obtained in (\ref{defpc}) is not of any use unless
   we can fix  $(c_1,c_2)$. For static configurations, we can convert the extra term to an integral over time 
   and a two-dimensional spatial surface. This is  true for any system, but in general, the result will not have 
   any simple form and will involve an undetermined
   range of integration over time coordinate.
   But in the case of gravity, two natural features conspire together to give an elegant form to this
   surface term. First is the fact that Rindler frame has a periodicity in Euclidean time and the range
   of integration over the time coordinate is naturally restricted to the interval $(0,\beta) = (0,2\pi/a)$.
   The second is the fact that the 
   surviving term in the 
   integrand $P^c$ is linear in the acceleration $a$ thereby neatly
   canceling with the $(1/a)$ factor arising from time integration. 
   [I will discuss these features more in section (\ref{EHaction}).]
   
    Given the form of $P^c$ we need to solve the 
   equation   
    \begin{equation}
 \left({\partial \sqrt{-g}L\over\partial g_{ab,c}}g_{ab}\right)=
 P^c= {1\over 4\pi {\cal A}_P} \left(2g^{cb}\partial_b\sqrt{-g} + \sqrt{-g}\partial_bg^{cb}\right)
 \label{dseq}
\end{equation}
to obtain 
the first order Lagrangian density.
    It is straightforward to show \cite{tpdlb} that this equation is satisfied by the Lagrangian
\begin{equation}
\sqrt{-g}L  = 
 {1\over 4\pi {\cal A}_P} \left(\sqrt{-g} \, g^{ik} \left(\Gamma^m_{i\ell}\Gamma^\ell_{km} -
\Gamma^\ell_{ik} \Gamma^m_{\ell m}\right)\right).
\label{ds}
\end{equation}  
    This is the second surprise. The Lagrangian which we have obtained is precisely
    the first order Dirac-Schrodinger Lagrangian for gravity (usually called the $\Gamma^2$
    Lagrangian).
    Note that we have obtained it without introducing the curvature tensor anywhere in the picture.
    Once again, this is unlikely to be a mere accident. 
    
    Given the two pieces, the final second order Lagrangian follows from our equation (\ref{aeh})
    and is, of course, the standard Einstein-Hilbert Lagrangian.    
   \begin{equation}
  \sqrt{-g} L_{grav}=\sqrt{-g}L - {\partial P^c\over\partial x^c} =  \left({1\over 4\pi {\cal A}_P}\right)R\sqrt{-g}.
      \label{lgrav}
       \end{equation}
    Thus our full second order Lagrangian  {\it turns out} to be the standard 
Einstein-Hilbert Lagrangian.
We have obtained this result
by just postulating that the surface term in the action should be proportional to the entropy per unit area.
This postulate uniquely determines the gravitational action principle and gives rise to a generally covariant
action.  The surface terms dictate the form of the Einstein Lagrangian in the bulk. 
The idea that surface areas  encode  bits of information per quantum of area  allows one to determine the nature of gravitational interaction on the bulk, which is an interesting realization of the holographic principle. 

I stress the fact that there is a very peculiar identity connecting the $\Gamma^2$ Lagrangian $L$ and the
Einstein-Hilbert Lagrangian $L_{grav}$, encoded in equation (\ref{lgrav}). This relation, which is purely a differential geometric identity, can be stated through the equations:
 \begin{equation}
   L_{grav} =L-\nabla_c\left[ g_{ab}
     {\partial  L \over \partial(\partial_c g_{ab})}
    \right]; \quad L=L_{grav}-\nabla_c\left[ \Gamma^j_{ab}
     {\partial  L_{grav} \over \partial(\partial_c \Gamma^j_{ab})}
    \right]
    \label{lageh}
   \end{equation}   
  This  relationship between the three terms defies any simple explanation in conventional
   approaches to gravity but arises very naturally in the approach presented here. 

The solution to (\ref{dseq}) obtained in (\ref{ds}) is not unique. However, self consistency requires that the final equations of motion for gravity must admit the line element in  (\ref{rindmetric}) as a solution.  It can be shown. by fairly detailed algebra, that this condition makes the Lagrangian in (\ref{dseq}) to be the only solution.
In particular, since we are demanding the flat spacetime to be a solution to the field equations, the cosmological constant in the pure gravity sector must be zero. [This, of course, does not prevent a cosmological constant arising from the matter sector of the theory.] 

\section{\label{EHaction}Structure of Einstein-Hilbert action}

 The approach leads to new insights regarding the peculiar structure of Einstein-Hilbert and  the $(3+1)$ formalism of gravity. To discuss these features, it 
    is convenient to temporarily switch to the signature $(- + + +)$ so that the spatial
  metric is positive definite.
 We will foliate the spacetime by a series
of space like hyper-surfaces  $\Sigma$ with $u^i$ as normal; then $g^{ik}=h^{ik}-u^iu^k$ where $h^{ik}$ is the induced metric on $\Sigma $. 
From the covariant derivative $\nabla_i u_j$ of the normals to $\Sigma $, one can construct only  three vectors
$(u^j\nabla_j u^i, u^j\nabla^i u_j, u^i\nabla^j u_j)$ which are linear in covariant derivative operator. The first one is
the acceleration $a^i=u^j\nabla_j u^i$; the second identically vanishes since $u^j$ has unit norm; the third,
$u^i K$, is
proportional to the trace of the extrinsic curvature $K=-\nabla^j u_j$ of $\Sigma $. Thus 
$V^i$ in the surface term in equation (\ref{aeh}) must be  a linear combination of $u^i K$ and $a^i$.
  In fact, one can
  show that (see, e.g.  equation (21.88) of the first reference in \cite{trkanda})
  \begin{equation}
  R= {}^3{\cal R} + K_{ab}K^{ab} - K_a^a K^b_b- 2 \nabla_i (Ku^i + a^i)\equiv {\cal L} - 2 \nabla_i (Ku^i + a^i)
  \label{rinh}
  \end{equation}
  where ${\cal L}$ is the ADM Lagrangian.  To prove this, we begin with the relation
  \begin{equation}
  R= - R g_{ab}u^au^b = 2 (G_{ab}-R_{ab})u^au^b
  \label{rinhone}
  \end{equation}
  and rewrite the first term using the identity:
  \begin{equation}
  2 G_{ab}u^au^b = {}^3{\cal R} - K_{ab}K^{ab} 
  + K_a^a K^b_b
  \label{ginr}
  \end{equation}
  As for the second term in (\ref{rinhone}), we note that $R_{abcd}u^d = (\nabla_a \nabla_b u_c - \nabla_b \nabla_a u_c)$
  giving 
  \begin{eqnarray}
  R_{bd}u^bu^d &=& g^{ac} u^bu^d R_{abcd} =  (u^b\nabla_a \nabla_b u^a - u^b \nabla_b \nabla_a u^a)
  \nonumber \\
  &=&  \nabla_a(u^b \nabla_b u^a) - ( \nabla_au^b)( \nabla_b u^a)-\nabla_b(u^b \nabla_a u^a)
  +( \nabla_b u^b)^2
  \nonumber \\
  &=& \nabla_i(Ku^i +a^i) - K_{ab}K^{ab} + K_a^a K^b_b
  \label{rabcd}
  \end{eqnarray}
  Using (\ref{rabcd}) and  (\ref{ginr}) we can rewrite (\ref{rinhone}) in the form of (\ref{rinh}).
  
  Let us now use (\ref{rinh}) to integrate $(R/16\pi)$
   over a four volume ${\cal V}$ bounded by two space-like
   surfaces $\Sigma_1$ and $\Sigma_2$ (with normals $u^i$) and two time-like surfaces
   ${\cal S}_1$ and ${\cal S}_2$ (with normals $n^i$). 
    The induced metric  on the space-like surface $\Sigma$ is $h_{ab} =
   g_{ab} + u_au_b$ while the metric on the time-like surface ${\cal S}$ is $\gamma_{ab} = g_{ab}
   - n_an_b$.
    These two surfaces will intersect on a two-dimensional surface ${\cal Q}$  on which the metric
    is $\sigma_{ab} = h_{ab}  - n_an_b = g_{ab} + u_au_b - n_an_b$. 
 Integrating both sides of (\ref{rinh}) over ${\cal V}$ we now get 
  \begin{eqnarray}
 A_{\rm EH} &=& \frac{1}{16\pi}  \int_{\cal V} R\sqrt{-g} \, d^4 x 
    = \frac{1}{16\pi} \int_{\cal V} {\cal L} \sqrt{-g} \, d^4 x -
  \frac{1}{8\pi} \int_{\Sigma_1}^{\Sigma_2} K\sqrt{h} \, d^3 x \nonumber \\
    && \hskip 10em -
   \frac{1}{8\pi} \int_{{\cal S}_1} ^{{\cal S}_2}(a_in^i)
  \sqrt{\sigma} \, d^2 x\, N\,dt
  \label{bigeqn}
  \end{eqnarray}
  where $g_{00} = - N^2$.
  In a static spacetime with a horizon: (i) $K=0$ making the second term on the right hand side vanish.
  (ii) The integration over $t$ becomes multiplication by $\beta$. (iii) Further, as the surface
  ${\cal S}_1$ approaches the horizon, the quantity $N (a_in^i)$ tends to
  $(-\kappa)$ where $\kappa$ is the surface gravity of the horizon, which is constant over the horizon.
  Using $\beta \kappa=2\pi$, the last term gives, on the horizon, the 
  contribution
  \begin{equation}
  \frac{\kappa}{8\pi} \int_0^\beta dt\int d^2x\, \sqrt{\sigma} = \frac{1}{4} {\cal A}_H
  \end{equation}
  where ${\cal A}_H$ is the area of the horizon. In the Euclidean sector 
  the first term gives $\beta E$ where $E$ is the
  integral of the ADM Hamiltonian over the spatial volume. We thus 
  get the result 
  \begin{equation}
   A_{\rm EH}^{\rm Euclidian} = \frac{1}{4} {\cal A}_H- \beta E  =  (S - \beta E)
  \end{equation}
  which is the free energy.
 For any static spacetime geometry, having a periodicity $\beta$ in the Euclidean time, the  Euclidean   
gravitational action represents the free energy of the spacetime; the first order term
gives the Hamiltonian  and the surface term gives  the entropy.

More generally, the analysis  suggests a remarkably simple, thermodynamical, interpretation of semiclassical gravity. In any
static spacetime with a metric
\begin{equation}
ds^2 = N^2({\bf x}) dt^2 - \gamma_{\alpha\beta}({\bf x})dx^\alpha dx^\beta
\end{equation}
 we have $R = {}^3R+2 \nabla_i a^i$ where $a_i=(0,\partial_\alpha N/N)$ is the acceleration of ${\bf x} = $ constant world lines. Then, limiting the time integration to $(0,\beta)$, say, the Einstein-Hilbert action becomes
\begin{equation}
A = \frac{\beta}{16\pi}\int_{\mathcal{V}}d^3 x N\sqrt{\gamma}{}^3R+
 \frac{\beta}{8\pi} \int_{\partial\mathcal{V}} (a^\alpha n_\alpha) d^2\mathcal{S}\equiv\beta E-S
\end{equation}
where the first term is proportional to energy (in the sense of spatial integral
 of  ADM 
Hamiltonian) and the second term is proportional to entropy in the presence of horizon. The variation of this action --- which leads to Einstein's equation --- is equivalent to the thermodynamic identity. (This result is explored in detail for spherically symmetric spacetimes in \cite{tplongpap}).

The surfaces
    $\Sigma, {\cal S}$ as well as the two surface ${\cal Q}$ on which they intersect will have corresponding extrinsic
    curvatures $K_{ab}, \Theta_{ab}$ and $q_{ab}$. 
In the literature, it is conventional to write the Einstein-Hilbert action as a term having only the first derivatives, plus an integral over the trace of the extrinsic curvature of the bounding surfaces. 
It is easy to obtain this form using the foliation condition $n^iu_i=0$ between the surfaces and noting:
  \begin{equation}
  n_ia^i=n_iu^j\nabla_ju^i=-u^ju^i\nabla_jn_i=(g^{ij}-h^{ij})\nabla_jn_i=-(\Theta-q)
 \label{thetaink}
  \end{equation} 
  where $ \Theta\equiv\Theta^a_a$ and $ q\equiv q^a_a$ are the traces of the extrinsic curvature of the 2-surface when treated as embedded in the 4-dimensional or 3-dimensional enveloping manifolds.
  Using (\ref{thetaink}) to replace $(a_in^i$) in the last term of (\ref{bigeqn}), we 
  get the result   
  \begin{eqnarray}
  A_{\rm EH} &+& \frac{1}{8\pi} \int_{\Sigma_1}^{\Sigma_2} K\sqrt{h} \, d^3 x
   -\frac{1}{8\pi} \int_{{\cal S}_1} ^{{\cal S}_2}\Theta
  \sqrt{\sigma} \, d^2 x\, Ndt \nonumber \\
    &=& \frac{1}{16\pi} \int_{\cal V} {\cal L} \sqrt{-g} \, d^4 x  -
   \frac{1}{8\pi} \int_{{\cal S}_1} ^{{\cal S}_2}q
  \sqrt{\sigma} \, d^2 x\, Ndt
   \label{}
  \end{eqnarray}
  In the first term on the right hand side, the ADM Lagrangian ${\cal L}$ contains ${}^3{\cal R}$
  which in turn involves the second derivatives of the metric tensor.
  The second term on the right hand side removes these second derivatives making the 
right hand side equal to the quadratic $\Gamma^2$ action  for gravity.  On the left hand side, the second and third
  terms are the integrals of the extrinsic curvatures over the boundary surfaces
  which, when added to the  Einstein-Hilbert
   action gives the quadratic action without second derivatives. (This is the standard result
often used in the literature). Unfortunately, this form replaces the normal component of the acceleration $a^in_i$ 
in (\ref{bigeqn}) by $(\Theta -q)$ and combines $q$ with ${}^3{\cal R}$ to get the first order Lagrangian. In the process, 
the normal component of the acceleration disappears and we miss the nice interpretation of Einstein-Hilbert action as the free energy of spacetime.

 \section{Conclusions}
  
  The approach adopted here is a natural extension of the original
  philosophy of Einstein; viz., to use non inertial frames judiciously to understand the 
  behaviour of gravity. In the original approach, Einstein used the principle
  of equivalence which leads naturally to the description of gravity  in terms of the 
  metric tensor. Unfortunately, {\it classical} principle of equivalence cannot take us any further
  since it does not encode information about the curvature of spacetime. However, the true
  world is quantum mechanical and one would like to pursue the analogy between non inertial
  frames and gravitational field into the quantum domain. Here the local Rindler frame arises
  as the natural extension of the local inertial frame and the study of the thermodynamics of the 
  horizon shows a way of combining special relativity, quantum theory and physics in the 
  non inertial frame. I have shown that these components are adequate to determine the
  action functional for gravity and, in fact, leads  to the Einstein-Hilbert action.
  This is remarkable because
  we did not introduce the curvature of spacetime explicitly into the discussion and 
  --- in fact --- the analysis was done in a  Rindler frame which is just
  flat spacetime. The idea works because the action for gravity splits up into
  two natural parts {\it neither} of which is generally covariant but are related to each other by the remarkable
identity (\ref{lageh}) which --- as far as I know --- was not noticed before.
The sum of the
  two parts  is generally covariant but the expression for individual parts can be ascertained in the 
  local Rindler frame specifically because these parts are {\it not} generally covariant.

  The fundamental postulate we use is in equation (\ref{postulate}) and it does
  {\it not} refer to any horizon. To see how this comes about, consider any
  spatial plane, say the $y-z$ plane, in flat spacetime. It is always possible to 
  find a Rindler frame in the flat spacetime such that 
  the chosen surface acts as the horizon for some Rindler observer.
  In this sense, any plane in flat spacetime must have an entropy per
  unit area. Microscopically, I would expect this to arise because of the 
  entanglement over length scales of the order of $\sqrt{{\cal A}_P}$.   
  We have defined in (\ref{postulate}) 
  the  entropy per unit area rather than the total entropy in order
  to avoid having to deal with global nature of the surfaces (whether the surface is 
  compact, non compact etc.). 
This approach also provides a natural explanation as to why the
  gravitational coupling constant is positive. It is positive because entropy and area
  are positive quantities. 
  
  The result emphasizes the role of two dimensional surfaces in fundamental physics.
A two dimensional surface is the basic minimum one needs to produce
region of inaccessibility and thus entropy from lack of information. When one connects up
gravity with spacetime entropy it is is inevitable that the coupling constant for gravity
has the dimensions of area in natural units. 
The next step in such an approach will be to find the fundamental units by which
  spacetime areas are made of and provide a theoretical, quantum mechanical
  description for the same. This will lead to the proper quantum description of spacetime
  with Einstein action playing the role of the free energy in the thermodynamic limit of the spacetime.
  
\section{Acknowledgement}

 I thank Apoorva Patel and K. Subramanian for several useful discussions.
 I thank the organizers of the ``Fred Hoyle's Universe" conference (Cardiff, June, 2002) for inviting me
 to give the talk and providing local hospitality.

\end{article}

\begin{thebibliography}{99}

 
  \bibitem{birreltp} Birrell N.D and Davies P.C.W, {\it Quantum fields in curved space}, 
  (Cambridge University  Press, Cambridge, 1982).
  
  
\bibitem {tprealms} Padmanabhan T.,  
       Mod.Phys.Letts. A {\bf 17}, 923 (2002). [gr-qc/0202078]. 
 
     
\bibitem{bhentropy}
Bekenstein, J.D., {\it Phys. Rev. D} 
{\bf  7}, 2333 (1973);  Hawking S.W., {\it Comm.Math.Phys.}, {\bf 43}, 199 (1975)
Gerlach, U.H., {\it Phys. Rev. D} 
{\bf  15}, 1479 (1976);
t'Hooft, G., {\it Nucl. Phys.} 
{\bf  B256}, 727 (1985).
York, J., {\it Phys. Rev. D} 
{\bf  15}, 2929 (1985);
Zurek, W.H. and Thorne, K.S., {\it Phys. Rev. lett.} 
{\bf  54}, 2171 (1985); 
Bombelli. L et al., , {\it Phys. Rev. D} 
{\bf  34}, 3, 73 (1986);  For an earlier attempt, similar in spirit to the current paper,
see Jacobson, T.  Phys.Rev.Letts., {\bf 75}, 1260 (1995).


\bibitem{grf}
This idea was suggested in
T. Padmanabhan, {\it The Holography of gravity encoded in a relation between Entropy, Horizon Area and the Action for gravity} [Second Prize essay; Gravity Research Foundation Essay Contest, 2002] and elaborated
in  T. Padmanabhan ,  Mod.Phys.Letts. A , 17, 1147 (2002) [hep-th/0205278]. 


\bibitem{holo}
't Hooft, G {\it Dimensional Reduction in quantum gravity}, gr-qc/9310026; 
{\it The holographic principle}, hep-th/0003004; L. Susskind (1995) J.Math.Phys., {\bf 36}, 6377;
for a recent review, see
R. Bousso,, {\it The holographic principle}, hep-th/0203101.
       
      

\bibitem{tpdlb}
  Lynden-Bell D. and T. Padmanabhan, (1994), unpublished; 
Padmanabhan,T., {\it Cosmology and Astrophysics - through problems}
(Cambridge university press, 1996) p. 170; p. 325.

 \bibitem{tplongpap} Padmanabhan T., Class.Quan.Grav. (2002),  {\bf 19}, 5387;
       [gr-qc/0204019]. 
 


\bibitem{trkanda}
Miser, C.W., Thorne K.S., Wheeler, J.A. , {\it Gravitation}
(Freeman and co., 1973), p.520;
York,J.W (1987) in {\it Between Quantum and Cosmos}, eds W.H.Zurek et al (Princeton University
Press, Princeton, 1988), p.246.

 
    
    

\end{thebibliography}
\end{document}